\documentclass[prl,twocolumn]{revtex4}

\usepackage{amsmath}
\usepackage{graphicx}
\usepackage{hyperref}
\usepackage[T1]{fontenc}

\newcommand{\ket}[1]{| #1\rangle}
\newcommand{\bra}[1]{\langle #1|}

\newcommand{\LO}{\mathrm{LO}}
\renewcommand{\section}[1]{}
\renewcommand{\Re}{\mathrm{Re}}
\begin{document}

\title{Spectral density matrix of a single photon measured}
\bibliographystyle{apsrev}
\author{Wojciech Wasilewski}\email{wwasil@fizyka.umk.pl}
\author{Piotr Kolenderski}
\author{Robert Frankowski}
\address{Institute of Physics, Nicolaus Copernicus University,
Grudzi\k{a}dzka 5, 87-100 Toru{\'n}, Poland}

\begin{abstract}
We propose and demonstrate a method for measuring the spectral density matrix of a single photon pulse.
The method is based on registering Hong-Ou-Mandel interference between photon to be measured and a pair of
attenuated and suitably delayed laser pulses described by a known spectral amplitude. The density matrix is
retrieved from a two-dimensional interferogram of coincidence counts. The method has been implemented for a
type-I downconversion source, pumped by ultrashort laser pulses. The experimental results agree well with
a theoretical model which takes into account the temporal as well as spatial effects in the source.
\end{abstract}

\maketitle

\section{Introduction}
Development of single photon sources brings the promise of implementing novel quantum-enhanced technologies.
In many applications, including quantum computing based on linear optics \cite{QAppli} photon sources are
required not only to deliver single light quanta, but also to supply them in a well defined mode. This is a
necessary condition for quantum interference between independent sources \cite{RiedmattenPRA03} which is
required in the above schemes. Also observation of three and more photon interference effects puts stringent
requirements on the sources \cite{XiangPRL06}. Besides, characterizing single photons is also interesting
from the fundamental point of view. Historically photons were first described within the framework of quantum
field theory (see \cite{BirulaPhWF} for a review) but more recently it was pointed out that a photon wave
function can be introduced \cite{BirulaPhWF,SipePRA95}. This kind of description, generalized by the
introduction of the density matrix constructed out of projectors on the states with specific wavefunctions,
seems to be the most elegant and effective theoretical tool for developing quantum-enhanced technologies
\cite{RhodeQPH06}.

Up till now measurements of polarization \cite{WhitePRL99} and spatial density matrix \cite{SmithOL05} of a
single photon were reported. The temporal characteristics of single photons were assessed only by verifying
whether they interfere with other sources \cite{RarirtyPTRSA97,PittmanQPH04} or between themselves
\cite{HOM}. In this Letter we propose and demonstrate a method for complete characterization of the temporal
degree of freedom: a measurement of the spectral density matrix of a single photon. We show that the
two-dimensional map of coincidence counts recorded as a function of delays between an unknown photon and a pair
of weak reference pulses can be used to reconstruct the magnitude and the phase of the density matrix. We present
a measurement for a type-I spontaneous down-conversion process in a bulk $\beta$-barium borate (BBO) crystal,
and compare the results of the reconstruction with theoretical predictions.

\section{Method} \label{sec:measurement}

When a single photon is launched into a singlemode fiber, the situation is
significantly simplified since the spatial mode is well defined. If moreover polarization of the photons is
fixed, the only remaining degree of freedom is the spectral one. In this case a single photon component of
the field can be described by the following density operator:
\begin{equation}\label{eq:hat_rho}
\hat \rho=\iint d\omega d\omega' \rho(\omega,\omega')\hat a^\dagger(\omega) \ket{0}\bra{0}\hat
a(\omega')
\end{equation}
where $\hat a(\omega)$  is an operator annihilating photon of frequency $\omega$ in the fiber, while
$\rho(\omega,\omega')$ is a density matrix of a single photon given in the spectral domain.

Our method for measuring $\rho(\omega,\omega')$ is based on the Hong-Ou-Mandel interference effect between
the single photon to be characterized and a local oscillator (LO) pulse of known shape attenuated to a single
photon level. The visibility of the two-photon interference dip is proportional to the overlap between the modes of
interfering photons. Figuratively speaking, by modulating the local oscillator spectral amplitude and
measuring the dip we can examine the single photon wavefunction from many directions. For a suitable class of
LO pulses this suffices to retrieve the spectral density matrix of a single photon. In a broad sense, our
experiment is a single photon analog of the homodyne method for measuring quantum correlations within a light
pulse \cite{DoubleHomo}. Indeed the density matrix describes correlations within a single
photon pulse.

\begin{figure}[b]
    \begin{center}
        \includegraphics[scale=0.5]{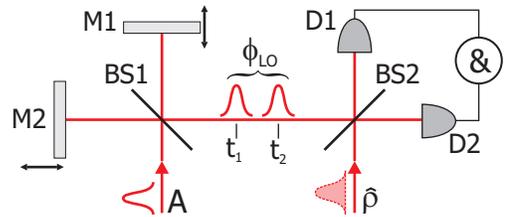}
        \caption{(Color online) Scheme of the experiment. A weak laser pulses described by a spectral
        amplitude $A(\omega)$ enter a Michelon interferometer, comprising mirrors M1 and M2 and a 50/50
        beamsplitter BS1, in which a double pulse of the local oscillator $\phi_\LO(\omega)$ is produced.
        Then the LO interferes with an unknown photon described by the density operator $\hat\rho$ on a 50/50
        beamsplitter BS2 and may give rise to a coincidence click of the detectors D1 and D2.}
        \label{fig:concept}
    \end{center}
\end{figure}
The method is presented in Fig.~\ref{fig:concept}. The first part is a Michelson interferometer
which serves as a LO pulse modulator. The second part is the beamsplitter on which the Hong-Ou-Mandel
interference occurs. When a master laser pulse described by a spectral amplitude function $A(\omega)$
enters the interferometer, it is split into two pulselets centered around delays $t_1$ and $t_2$ determined by
the length of the arms in the interferometer. If both pulselets overlap, they interfere on the beamsplitter
and depending on the precise value of their relative delay $t_2-t_1$ the input pulse energy is directed
alternatively towards the output of the interferometer or back to the laser. The normalized spectral amplitude of
local oscillator pulses prepared this way reads:
\begin{equation}\label{eq:phiLO}
\phi_\LO(\omega)= A(\omega)\frac{\exp(-i\omega t_1)+\exp(-i\omega t_2)}{\sqrt{2S(t_2-t_1)}}
\end{equation}
where $S(t_2-t_1)$ is a normalizing function equal to the probability that a single photon will pass through
the Michelson interferometer. The modulated LO pulses interfere with unknown single photons on a 50/50
beamsplitter BS2. If the master laser pulse contains a photon with probability $l$, while the source produces
photon described by spectral density matrix $\rho(\omega,\omega')$ with probability $f$ per pulse, we can
calculate the probability of registering coincidence with detectors of quantum efficiency $\eta$ to be
\cite{RhodeQPH06}:
\begin{equation}\label{eq:p_c}
p_c(t_1,t_2)=flS(t_2-t_1)\frac{\eta^2}{2}\left[1-Q(t_1,t_2)\right],
\end{equation}
where we have assumed that each of the input states contains at most one photon and $Q(t_1,t_2)=\iint d\omega
d\omega'\, \phi^*_\mathrm{LO}(\omega) \rho(\omega,\omega') \phi_\mathrm{LO}(\omega')$ is an overlap between
the local oscillator $\phi_\mathrm{LO}(\omega)$ and the density matrix $\rho(\omega,\omega')$. We can
partially evaluate $Q(t_1,t_2)$ using Eq.~\eqref{eq:phiLO}:
\begin{eqnarray}\label{eq:phirhophi_2pulses}
Q(t_1,t_2)&=&\frac{D(t_1)+D(t_2)+2\Re\tilde\rho(t_1,t_2)}{2S(t_2-t_1)}
\end{eqnarray}
where $\Re$ stands for real part, while
\begin{multline}
\tilde\rho(t_1,t_2)=\iint d\omega_1 d\omega_2\,e^{i\omega_1t_1-i\omega_2t_2}\\
A^*(\omega_1) \rho(\omega_1,\omega_2) A(\omega_2)
\end{multline}
is the inverse Fourier transform of a density matrix restriced to the spectral domain of the master laser
pulses and $D(t)=\tilde\rho(t,t)$ is the function describing the standard Hong-Ou-Mandel interference dip. Inserting
Eq.~\eqref{eq:phirhophi_2pulses} into Eq.~\eqref{eq:p_c} yields:
\begin{equation}\label{eq:rho=I-NC-T}
Np_c(t_1,t_2)=2S(t_2-t_1)-2\Re \tilde \rho(t_1,t_2)-D(t_1)-D(t_2)
\end{equation}
where $N$ is a normalization factor combining the probabilities of detecting photon from the master laser $l$ and
the source $f$, as well as detector efficiency $\eta$. At this point let us note that the density matrix
$\rho(\omega,\omega')$ is nonzero  only near some point in the frequency space $\omega=\omega'=\omega_0$ in
case of narrowband photons, where $\omega_0$ is the central frequency of the single photons which we will
assume to be equal to the central frequency of the master laser. Therefore $\Re \tilde \rho(t_1,t_2)$
oscillates like $\cos[\omega_0 (t_2-t_1)]$. Also $S(t_2-t_1)$ contains such oscillations. On the other
hand $D(t)$ is a slowly varying function of its argument. Therefore individual components of the right hand side of
Eq.~\eqref{eq:rho=I-NC-T} can be separated in the frequency domain. Let us apply the Fourier transform $\iint dt_1 dt_2
\exp(-i\omega_1t_1+i\omega_2t_2) \ldots$ to both sides of Eq. \eqref{eq:rho=I-NC-T} and rearrange terms:
\begin{multline}
A^*(\omega_1) \rho(\omega_1,\omega_2) A(\omega_2)
+A(-\omega_1) \rho^*(-\omega_1,-\omega_2) A^*(-\omega_2)\\
=\delta(\omega_1-\omega_2)|A(\omega_1+\omega_2)|^2 -N\tilde p_c(\omega_1,\omega_2) + \ldots
\end{multline}
where by dots we have denoted Fourier transform of $D(t_1)$ and $D(t_2)$, while $\tilde
p_c(\omega_1,\omega_2)$ is the Fourier transform of coincidence counts. Additionally we used the fact that
$|A(\omega)|^2$ is the Fourier transform of $S(t)$.

In the experiment we can directly measure $p_c(t_1,t_2)$ as a function of $t_1$ and $t_2$. In the case of
narrowband single photons retrieving of the spectral density matrix runs the following way: first
Fourier transform of $p_c(t_1,t_2)$ is computed. A region in the frequency space where contribution from
$S(t_2-t_1)$ and $\tilde\rho(t_1,t_2)$ lies is separated. Next, a large contribution from $S(t_2-t_1)$ is
calculated and subsequently subtracted by measuring $p_c(t_1,t_2)$ for large $t_1$ and $t_2$ where
$\tilde\rho(t_1,t_2)$ is zero. This way we obtain the Fourier transform of $\tilde\rho(t_1,t_2)$, which equals
$A^*(\omega_1) \rho(\omega_1,\omega_2) A(\omega_2)$. It is divided by spectral amplitude of master laser
pulses $A(\omega)$, which is measured separately and finally $\rho(\omega_1,\omega_2)$ is found. Note that
the last step is well defined only for single photons of bandwidth narrower than that of the master laser.

\begin{figure}
    \begin{center}
        \includegraphics{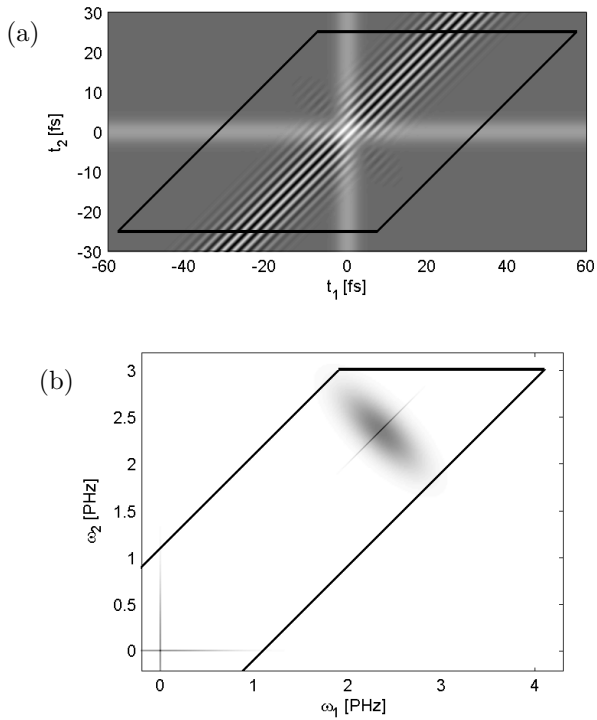}
        \caption{(a) A typical coincidence interferogram as a function of delays $t_1$ and $t_2$
        and (b) its Fourier transform. The black parallelogram in (a) defines the scan range, which with suitable sampling
        density yields the region of interest in the frequency domain, outlined in (b).}
        \label{fig:coinc}
    \end{center}
\end{figure}
To illustrate the above dry formulas a typical coincidence pattern and its Fourier transform are depicted in
Fig.~\ref{fig:coinc}. The diagonal fringes in Fig.~\ref{fig:coinc}(a) come from $S(t_2-t_1)$, the vertical
and horizontal stripes from $D(t_1)$ and $D(t_2)$ while the most interesting term $\tilde \rho(t_1,t_2)$ contributes
only in the very center of the picture. It is easier identified in Fourier transform plot in
Fig.~\ref{fig:coinc}(b) where a diagonal cloud corresponds to $A^*(\omega_1) \rho(\omega_1,\omega_2) A(\omega_2)$,
$D(t_1)$ and $D(t_2)$ contribute a cross arround zero frequency while Fourier transform of
$S(t_2-t_1)$ is seen as a ridge along $\omega_1=\omega_2$.

In the above derivations we have assumed interference between single photons. In our experiment, in fact we
interfere a weak coherent state with a multimode thermal state. Whereas the visibility of such interference
can be exactly calculated using the semiclassical theory \cite{RarirtyPTRSA97}, two photon terms of the
coherent state and the thermal state contribute only towards a constant background of coincidence counts.
However, the shape of the interference pattern does characterize the single photon component of the signal
field. Its density matrix can still be retrieved in the way described above.

\section{Experiment and Results}\label{sec:experiment}

\begin{figure}[b]
    \begin{center}
        \includegraphics[scale=0.6]{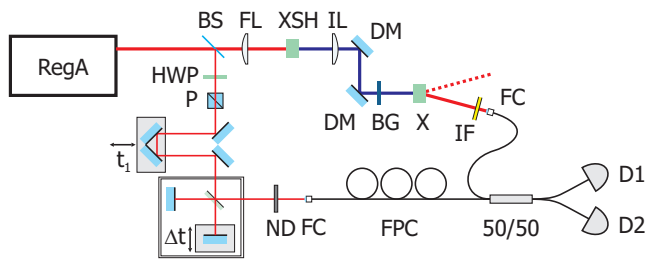}
        \caption{(Color online) The experimental setup. BS, beamsplitter; FL, focusing lens; XSH, BBO crystal for
        generation of the second harmonic; IL, imaging lens; DM dichroic mirrors; BG, blue glass filter;
        X, downconversion crystal; IF, interference filter; FC, fiber coupling stage, HWP,
        half waveplate; P, polarizer; ND, neutral density filter; FPC, fiber polarization controler; D1 and D2,
        single photon counting modules.}
        \label{fig:setup}
    \end{center}
\end{figure}

Our experimental setup is depicted in the Fig.~\ref{fig:setup}. The master laser (RegA 9000 from Coherent)
produces a train of 165~fs FWHM long pulses at a 300kHz repetition rate centered at 774~nm, of which we use
300~mW average power. Most of the energy goes to the second harmonic generator, based on a 1~mm thick BBO crystal
cut for type-I process. Ultraviolet pulses produced this way have 1.3nm bandwidth and 30~mW average power.
They are filtered out of fundamental using a pair of dichroic mirrors DM and a color glass filter BG (Shott
BG39), and imaged using 20cm focal length lens IL on a downcoversion crystal X, where they form a spot
measured to be 155~$\mu$m in diameter. The crystal X is a 1~mm thick BBO crystal cut a $29.7^\circ$ to the
optic axis, and oriented for maximum source intensity. A portion of down-converted light  propagating at an
angle of $2.5^\circ$ to the pump beam passes through a 10nm interference filter centered at 774.5nm and is
coupled into a single mode fiber. This defines the spatial mode in which the down-conversion is observed
\cite{Dragan2004}. About 4\% of energy of master laser pulses is reflected towards an LO preparation arm. The
pulses first go through a half wave plate HWP and a polarizer P allowing for fine control of the energy. Then
they are delayed in a computer-controlled delay line. Next the pulses enter a Michelson interferometer in
which one mirror moves on a precision computer controlled stage allowing for generation of double pulses with
a well-defined temporal separation $\Delta t=t_2-t_1$. Finally the double pulses are attenuated to contain less
than 0.1 photon on average and coupled to a single mode fiber, where their polarization is adjusted using
a fiber polarization controller FPC to match the polarization of the photon coming from the downconversion source.
Both the downconversion and local oscillator photons interfere in a 50/50 singlemode fiber coupler and are
detected using single photon counting modules SPCM (PerkinElmer SPCM-AQR-14-FC) connected to fast coincidence
counting electronics (suitably programmed Virtex4 protype board ML403 from Xilinx) detecting events in
coincidence with master laser pulses.

\begin{figure}
    \begin{center}
        \includegraphics[scale=0.75]{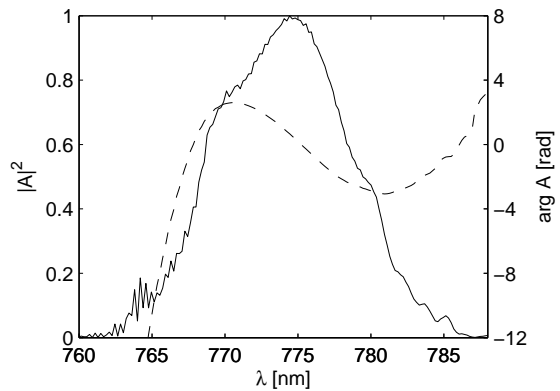}
        \caption{Spectral intensity  $|A(\omega)|^2$  (solid line) and phase $\arg A(\omega)$ (dashed line) of the master
        laser pulse retrieved using FROG.} \label{fig:ALO}
    \end{center}
\end{figure}
For calculating the actual density matrix of unknown single photons, a characterization of master laser pulses
was necessary. This was accomplished using Frequency-Resolved Optical Gating (FROG) technique
\cite{FROGbook}. Retrieved spectral intensity  $|A(\omega)|^2$ and phase $\arg A(\omega)$ are plotted in
Fig.~\ref{fig:ALO}. Precise knowledge of the master laser pulses allowed us also to calculate the second
harmonic pulse using perturbative approach.

\begin{figure}
    \begin{center}
    \includegraphics[scale=0.7]{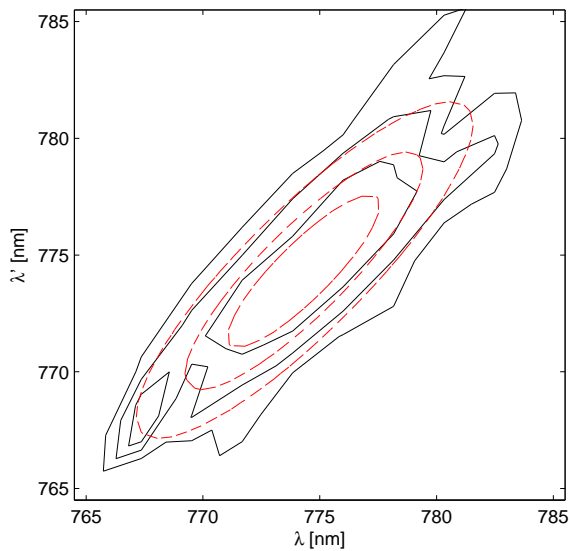}
   \caption{(Color online) Contour plot of the measured spectral density matrix $\rho(\omega,\omega')$
   as a function of wavelengths $\lambda=2\pi c/\omega$ (solid lines) and a theoretical prediction
   for this quantity (dashed red lines). The contour were drawn at 0.75, 0.5 and 0.25 of the maximum values,
   the outermost encircles 7$\times$5 experimental data points.}
   \label{fig:result}
    \end{center}
\end{figure}

The complete measurement consisted in a series of 6 scans of a rectangular grid depicted in
Fig.~\ref{fig:coinc}(a) spanned by 4000$\times$25 points, where the latter number refers to the direction
along the fringes. The corresponding mesh was 0.233~fs $\times$ 66~fs, and coincidences were counted for
80~ms at each point. The reconstructed spectral density matrix of a single photon is depicted in
Fig.~\ref{fig:result}. We compare it with theoretical calculations plotted with dashed lines in the same
figure. The theoretical model used in these calculations assumed the exact phase matching function of the
nonlinear crystal and the ultraviolet pump pulse shape computed from the measured $A(\omega)$. The transverse
components of the wave vectors for the pump and down-converted beams were treated in the paraxial
approximation. The spectral density matrix was calculated for coherent superpositions of plane-wave
components of the down-conversion light that add up to localized spatial modes defined by the collecting
optics and single-mode fiber. The other photon from the source, which remains undetected, was traced out
assuming that it can propagate at any direction and have any frequency that is consistent with the
conservation of energy and perpendicular momentum in the downconversion crystal. As seen in
Fig.~\ref{fig:result} the theoretical calculations predict more pronounced correlations i.e. smaller width
along the antidiagonal $\rho(\omega_0+\delta,\omega_0-\delta)$ than was actually measured. We attribute this
discrepancy to a difference between the actual ultraviolet pump pulse shape and the one calculated from $A(\omega)$.
Also the tips of the density matrix are measured with reduced accuracy, since in that regions the raw
experimental result is divided by a relatively small master laser spectral intensity $A(\omega)$ which
amplifies errors. The theoretical model predicts the phase of the density matrix to be smaller than $\pi/30$
in the region bounded by the contour at 0.25 maximum. The measured phase in this region is smaller than $\pi/10$
and varies randomly from point to point.

\section{Conclusions}

In summary, we proposed and demonstrated a method for measuring the spectral density matrix of a single photon
component of the electromagnetic field in a singlemode fiber. The method is based on two photon interference
and is thus limited to the spectral range where known reference pulses are available, however it allows for
retrieving both amplitude as well as phase of the density matrix. We have applied this method to a
downconversion-based source of single photons and found that measured density matrix agrees with theoretical
predictions.

We acknowledge insightful discussions with K. Banaszek. R.~F. acknowledges discussions with M.~Zieli{\'n}ski,
M.~Kowalski, S.~Grzelak and D.~Chaberski. This work has been supported by the Polish budget funds for
scientific research projects in years 2006-2008, the European Commission under the Integrated Project Qubit
Applications (QAP) funded by the IST directorate as Contract Number 015848 and AFOSR under grant number
FA8655-06-1-3062. W.W. gratefully acknowledges the support of the Foundation for Polish Science (FNP) during
this work. It has been carried out in the National Laboratory for Atomic, Molecular, and Optical Physics in
Toru\'{n}, Poland.

\newcommand{\urlprefix}{ }
\renewcommand{\url}[1]{}


\end{document}